\documentclass[a4paper,11pt]{article}
\usepackage[main=english,french]{babel}
\usepackage{hyphenat}                      
\usepackage{hyperref}
\hypersetup{colorlinks=true, citecolor=MidnightBlue,
            linkcolor=MidnightBlue, urlcolor=MidnightBlue, linktocpage=true}
\usepackage{amsmath,amsthm,amssymb,amsfonts,physics,stmaryrd} 
\usepackage{empheq}  % for boxing equations
\usepackage{cancel}
\usepackage{dsfont,cancel,slashed,bbold,mathrsfs}
\usepackage{graphicx,wrapfig,caption,subcaption,setspace}
\usepackage[export]{adjustbox}
\usepackage{indentfirst,enumitem,lastpage,sectsty}
\usepackage{fancyhdr,titlesec,anyfontsize}
\usepackage[dvipsnames]{xcolor}
\usepackage[margin=2.5cm]{geometry}
\usepackage{xcolor}
\usepackage{cite}
\theoremstyle{remark}

\renewcommand{\dd}[2][]{\mathop{}\!\mathrm{d}^{#1}#2 \, }
\newcommand{\sign}{\operatorname{sgn}}

\title{\LARGE\bfseries
Probabilistic Causality from Graviton Fluctuations}
\author{Giordano Cintia\footnotemark[1]
,  Federico Piazza\footnotemark[2]
, and Samuel Ramos\footnotemark[3]
\\[2mm]
\small\textit{Aix Marseille Univ, Universit\'e de Toulon, CNRS, CPT, Marseille, France} 
}

\date{}
\begin{document}
\maketitle
\begingroup
\renewcommand{\thefootnote}{}
\footnotetext{$\!\!\!^*$\href{mailto:}{giordano.cintia@univ-amu.fr};\, $^\dagger$\href{mailto:}{piazza@cpt.univ-mrs.fr}; \,$^\ddagger$ \!\href{mailto:}{samuel.ramos@cpt.univ-mrs.fr}}
\endgroup
\vspace{-4 mm}
\begin{abstract}
 We compute the commutator of a scalar field minimally coupled to gravity at leading order in $G_N$. The commutator is operator-valued, with terms involving derivatives of Dirac deltas supported on the Minkowski light cone. When evaluated on classical/coherent graviton states, these terms ``bend" the support of the commutator in precisely the way required to recover standard causality on a classical curved
 spacetime. However, these terms are also associated with a variance and are thus a source of uncertainty in the causal relations between events. We quantify this effect for a thermal state of gravitons at temperature $T$. In the large-time limit, the uncertainty of the lightcone position is given by
 $$
\text{Var}(\delta x) =\frac{4G_NTt}{3}\, ,
$$
where $\delta x$ is the displacement from the classical lightcone at $|\vec x| = t$. 
We obtain this result by subtracting a universal vacuum contribution that is logarithmically UV divergent and subleading at late times. 
\end{abstract}

\section{Introduction and summary}
Commutators of local observables are a direct probe of causality, because they determine whether measurements made at separate events can affect each other. 
When fields are quantized on a classical background spacetime, such commutators vanish at spacelike separation---this is the principle of \emph{microcausality} \cite{Peskin:1995ev,Dubovsky:2007ac}. In this limit, the analytic properties of the commutator, or equivalently of the Green functions, encode detailed information about the causal structure of the theory~\cite{Creminelli:2022onn,Creminelli:2023kze,Creminelli:2024lhd, Heller:2022ejw, Hui:2023pxc, Hui:2025aja}. Things get more subtle when gravity is taken to be dynamical. 
Strictly speaking, local field operators are no longer gauge invariant and therefore are not observables.  Challenging the principle of microcausality in this context can prove misleading---this is why one typically resorts to verifying subluminal propagation directly at spacetime asymptotics while ignoring what happens in the bulk (e.g.~\cite{Gao:2000ga,Camanho:2014apa,deRham:2021bll,Bellazzini:2021shn,Bucciotti:2026svg} and related references). 

Of course, one can still make ``localized" measurements in the bulk of a fluctuating spacetime.  It is just that some coordinate-fixing prescription must be attached to the corresponding operators to specify, in a gauge-invariant manner, the position at which the measurement is performed. This results in mildly non-local observables. For example, in asymptotically AdS, a location in the bulk is often defined by sending in a geodesic from a point of the boundary.   The resulting operators are dressed by a gravitational Wilson line.\footnote{On gravitational dressing we refer the reader to the classic papers by Donnelly and Giddings~\cite{Donnelly:2015hta,Donnelly:2016rvo} (see also~\cite{Giddings:2019wmj}). For specific applications to the AdS/CFT correspondence, see e.g.~\cite{Papadodimas:2015jra,Almheiri:2017fbd,Blommaert:2019hjr} and references therein.}  

When dressed along timelike geodesics, these operators are particularly meaningful because they represent the observables available to a set of physical observers (see Sec.~\ref{frame}). Their commutators are a direct probe of the causal relations experienced by such observers. It is interesting to ask how such relations are affected in generic gravitational states.
One expects that the quantum fluctuations of the metric can make the lightcone structure uncertain and the causal relations probabilistic~\cite{Hartle:1992as,Hardy:2006uc,Zych:2017tau}. The purpose of this paper is to estimate this uncertainty in a concrete setup. 

In Sec.~\ref{sect:3} and App.~\ref{A}, we compute the commutator of a test scalar field $\phi$ minimally coupled to gravity.  We work in perturbation theory around Minkowski space. To lowest non-trivial order in $G_N$, we find that the relevant terms in this commutator are
\begin{equation} \label{retarintro}
2 \pi i \, [\phi(x), \phi(0)] = \delta(t^2 - \vec x^{\, 2}) + \kappa\,  t \, {\cal O}( x) \delta'(t^2 - \vec x^{\, 2}) \, ,
\end{equation}
where $\kappa = \sqrt{8\pi G_N}$, ${\cal O}( x)$ is a non-local operator linear in the metric defined below in eq.~\eqref{o}. Also, positive times $t>0$ have been assumed in the above. 
There is a sense in which one can interpret the RHS as a Taylor expansion of a Dirac delta with a shifted support, \emph{i.e.} $\simeq \ \delta(t^2 - \vec x^{\, 2} +  \kappa \,t\,  {\cal O}( x)  )$~\cite{Cintia:2025fzn}. In order to check this, one should integrate these distributions over a smooth tempered function. Indeed, when applying linear response theory to the retarded propagator~\eqref{retarintro}, such a smooth function is simply the source $J(x)$, as shown in Sec.~\ref{sec_linear}. 

Under the hypotheses discussed in that section, the support of the commutator of a scalar in the presence of gravity is effectively displaced with respect to the standard Minkowski lightcone by an amount proportional to the non-local operator ${\cal O}( x)$. The latter has a mean and a variance. On semiclassical gravitational states, the graviton field acquires a background value. In this case, the mean of ${\cal O}( x)$ reproduces the expected lightcone of the corresponding curved spacetime. The variance of ${\cal O}( x)$, on the other hand, measures how uncertain causality becomes as a consequence of graviton fluctuations. 

A relatively well-controlled setting in which to apply this formalism is a thermal state of gravitons at temperature $T$. The variance, calculated in App.~\ref{B} and discussed in Sec.~\ref{sec_5}, separates into vacuum and thermal contributions. In terms of the displacement $\delta x$ from the Minkowski lightcone, these take the form
\begin{equation} \label{varint}
  \text{Var}(\delta x)_{\rm vac}
    \simeq G_N \log\left(\frac{t}{R}\right)\,,
    \qquad
   \text{Var}(\delta x)_{\rm therm}
    = \frac{4 G_N T\,t}{3}\,,
\end{equation}
where $t$ is the elapsed time since emission and $R$ is a UV regulator with dimensions of length. The situation is reminiscent of entanglement entropy, where a finite piece induced by a nonzero temperature adds to the UV-divergent vacuum contribution~\cite{Holzhey:1994we,Calabrese:2004eu}. The latter reflects the short-distance behavior of all Hadamard states and diverges as $R\to 0$, whereas the thermal contribution is finite and grows linearly with both $T$ and $t$. Although the thermal behavior dominates at late times, one could still wonder about the physical meaning of the subleading vacuum term.  Should it be subtracted, or does it represent an observable contribution to the lightcone spread? As argued in Sec.~\ref{sec_5}, we tend to favor the latter interpretation, with $R$ physically representing the finite size of the source emitting the signal.

We conclude the paper by discussing in Sec.~\ref{comments} a few points of interest, among which a possible application to black holes. An approximately thermal graviton flux should inhabit the region close to a black hole as a consequence of Hawking radiation.
 This is also a situation in which the spreading of the lightcone may become physically relevant. Unlike an ordinary expanding light cone in Minkowski, the horizon of a black hole is a null surface whose size is maintained fixed. If the accumulated spread becomes comparable to the horizon size before evaporation is complete, this could have implications \emph{e.g.} for the black hole information problem. A straightforward, albeit heuristic, application of our formulas suggests that this may indeed be the case.

\paragraph{\textit{Notations and conventions}}
 We work in natural units $\hbar=c=1$, the only dimensional coupling being $\kappa^2=8\pi G_N$. We use the usual notation $\eta$ for the Minkowski metric with signature $(-+++)$. We omit hats on operators, as their operator nature is hopefully clear from the context. We often split the temporal and spatial parts of spacetime events. Spacetime events are denoted by $x$, $y$, etc., or by explicitly splitting their temporal and spatial parts, \textit{i.e.} $(x^0,\vec{x})$, $(y^0,\vec{y})$, etc. 

\section{Choice of frame} \label{frame}
 In the presence of dynamical gravity, different coordinate fixings, \emph{i.e.} different dressing procedures, effectively correspond to different definitions of an event. 
  This is, with a slogan, the \emph{relativity of the event}~\cite{Nitti:2024iyj} (see also~\cite{Giacomini:2017zju,Goeller:2022rsx,Kabel:2024lzr}).  As a consequence, the commutator of a massless field might have support on a sharp lightcone in some coordinates and exhibit uncertainty in a different set of coordinates.  In the simplest possible terms, this can be shown with a mini-superspace example. The metric element can be written in conformal time $\tau$, or in proper time $t$,
  \begin{equation}
ds^2 \ = \ a^2(\tau)(- d\tau^2 + d {\vec x}^{\, 2}) \ = \ - dt^2 + a^2(t)d {\vec x}^{\, 2}\, .
\end{equation}
In conformal time $\tau$, light rays satisfy $d |\vec x| = d\tau$, so the lightcone remains sharp regardless of how strongly the scale factor $a$ fluctuates.  In the $t$-coordinate, light rays instead obey $d |\vec x| = dt/a(t) $---fluctuations of $a$ now do induce an uncertainty of the lightcone. Of course, on a classical spacetime, switching from $\tau$ to $t$ is mere relabeling. But when $a$ is dynamical, switching to $t$ changes the nature of causal relations from sharp to probabilistic. 

While all coordinate choices are equivalent in principle, physical measurements correspond to a particular type of dressing, in which positions are tied to observers worldlines and time is the proper time along them.
As free-falling observers in the universe, we maintain continuous causal contact with nearby galaxies/observers. When we detect, for instance, a gamma-ray burst, we record the event at a definite value of our own proper time---the only notion of time directly accessible from our geodesic. The source event itself, such as the collapse of the progenitor star, is attached to another distant geodesic and triggered by the local physics there, \textit{i.e.}, by a definite proper time along that worldline.
Decoding causal relations in any other frame  is possible in principle, but very difficult in practice. 
In the example above, accessing the $\tau$ coordinate would require complete knowledge of the (quantum) state of the scale factor $a$.

For this reason, in this paper, we adopt coordinates that correspond to geodesic/proper time dressing.\footnote{Alternative interesting choices include \emph{Rindler-like} coordinates. In this case, the observables are attached to worldlines of constant acceleration. We postpone this study to future work.} To this aim, we consider metrics in the \emph{synchronous gauge} 
\textit{i.e.} with lapse $N = 1$  and shift $N_i =0$,
\begin{equation} \label{metric}
ds^2 = -dt^2 + g_{ij} dx^i dx^j \, .
\end{equation}
Curves with $x^i$ = const are indeed timelike geodesics in this metric and $t$ is their proper time. We work around Minkowski spacetime, with
\begin{equation}
g_{ij} = \delta_{ij} + \kappa h_{ij}\, .
\end{equation} 
Let us now consider a light ray passing through the origin in the spatial direction defined by the unit vector $\hat x$. %While finding its trajectory implies solving a geodesic equation, its longitudinal displacement (\textit{i.e.} along $\hat x$) can be  obtained directly from~\eqref{3}. 
To first order in $\kappa$ one finds that this is the sum of two contributions, 
\begin{equation} \label{3}
x^k(t) = \left(t - \frac{\kappa}{2}\int_0^t dt' \, h_{ij}(t', t' \hat x \, ) \, \hat x^i \hat x^j\right)\hat x^k\,+\, \kappa\, \delta x^k_T(t).
\end{equation}
The first term in parenthesis represents the longitudinal part of the geodesic---the one along $\hat x$.  The integral is taken along the Minkowski light ray, which is the zeroth-order solution to the null geodesic equation. This displacement is responsible for the deformation of the lightcone. The transverse displacement, $\delta x_T$, is obtained by solving the geodesic equation. However, it occurs along the surface of the lightcone itself and can therefore be ignored in the present analysis as it does not impact causality.

\section{The operator-valued  commutator}
\label{sect:3}
 
Equation~\eqref{3} displays the expected deviation from Minkowski causality due to the presence of a classical gravitational field $h_{ij}(x)$.  How do commutators of test fields ``know" of such a deviation in a quantum field theory? 
In a massless free theory the commutator is a $c$-valued function with support on the lightcone,
\begin{equation}\label{free}
[\phi_0(x), \phi_0(0)] \ \equiv \ i \Delta(x) \ = \ \frac{\sign(t)}{2 \pi i} \delta(t^2 - \vec x^{\, 2})\, , 
\end{equation}
where $t \equiv x^0$ has been defined for notational simplicity. 
In order to see the effects of gravity, let us consider the same scalar field $\phi$ coupled to Einstein gravity,
\begin{equation}
     S =  \int \dd[4] x \sqrt{-g} \, \left( \frac{1}{2\kappa^2}R -\frac{1}{2} g^{\mu\nu}\partial_\mu \phi \partial_\nu \phi\right).
    \label{action}
\end{equation}
We expand the metric around Minkowski space. At the lowest order in $\kappa$, the constraint equations in the synchronous gauge are solved by the \emph{transverse traceless} (TT) \emph{condition} $h_{ii} = \partial_i h_{ij}=0$, and hence coincide with the \emph{TT}-gauge. The Lagrangian to order $\kappa$ reads
\begin{align} \label{theory}
   {\cal L} =   
    -\frac{1}{8} \partial_\rho h_{ij} \partial^\rho h_{ij}-\frac12 \partial_\mu \phi \partial^\mu \phi +\frac{\kappa}{2} \, h_{ij} \, \partial_i \phi \partial_j \phi \, .
    \end{align}
When interactions are considered, the $\phi$-commutator is \emph{operator-valued} and can be computed in perturbation theory (e.g.~\cite{Cintia:2025fzn}). One way of doing so is to express the evolution operator in interaction picture in terms of time-ordered products of free fields. This gives
\begin{flalign}
    [\phi(x),\phi(y)]=i\Delta(x-y)+i\!\int_{t_*}^{t_x} \!\dd[4]z\,\big[\left[\mathcal{H}_I(z),\phi_0(x)\right],\phi_0(y) \big] -i \!\int_{t_*}^{t_y} \! \dd[4] z \, \big[\left[\mathcal{H}_I(z),\phi_0(y)\right],\phi_0(x)\big],
\end{flalign}
where the interaction Hamiltonian density is obtained from~\eqref{theory},   $ \mathcal H_I = - \frac{\kappa}{2} h_{ij}\partial_i \phi_0 \partial_j \phi_0$,
and $t_*$ is the reference time at which the Heisenberg- and interaction picture-operators coincide. It is easy to see that the dependence on $t_*$ cancels in the calculation of the commutator. 
In fact, by repeatedly applying~\eqref{free} one obtains 
\begin{align}
     [\phi(x),\phi(y)] 
     = i\Delta(x-y)+i\kappa\int^{t_x}_{t_y}\dd[4] z\, h^{ij}(z)\partial_{z^i}\Delta(z-x)\partial_{z^j}\Delta(z-y)\, .
     \label{eq:comm-intpic}
\end{align}
\emph{To ease the notation, in the main text we assume $t>0$ from now on,} which is equivalent to studying the \emph{retarded} propagator.
After some massaging of the above integral, one obtains a combination of delta and delta-derivatives supported on the lightcone\footnote{{A related result was recently obtained in Ref.~\cite{Allic:052026} by computing expectation values of a dressed commutator on a coherent state of gravitons.}}
%, whom we met at the conference Relativity Reframed: Quantum Reference Frames and Gravity, organized by the Perimeter Institute.} 
and Heaviside step functions $\theta$ supported \emph{inside} the lightcone (see App.~\ref{A} for details) 
\begin{align}  \nonumber
2 \pi i \, [\phi(x), \phi(0)] & \ =\ \delta(t^2 - \vec x^{\, 2}) \\ & \ \, + \, \kappa\,  \Big( t \, {\cal O}( x) \delta'(t^2 - \vec x^{\, 2}) + {\cal O}_1( x) \delta(t^2 - \vec x^{\, 2}) + {\cal O}_2( x) \theta(t^2 - \vec x^{\, 2}) \Big)\, . \label{structure2}
\end{align}
Higher orders in $\kappa$ contain various combinations of the same ingredients and higher delta-derivatives.

 The crucial operator here is $\mathcal{O}(x)$, discussed in the next section. The expressions of ${\cal O}_1( x)$ and ${\cal O}_2( x)$ are given in App.~\ref{A}.
The term proportional to the Dirac delta, ${\cal O}_1( x)$,  provides a modulation of the field's intensity along the lightcone.  
The last operator, proportional to the theta function, fills the inside of the lightcone. Once evaluated on some gravitational state, this represents the expected Hadamard tail that correlators exhibit on curved spacetime (see~\cite{Chavda:2025ijz} for a recent discussion). This operator is not particularly interesting for causality and we will thus ignore it from now on. 

Clearly, the above equation is distributional and should be interpreted with some care. Strictly speaking, this would require first evaluating the expectation values of the operators on a given state, and only afterwards taking the corresponding weak limit. The subtlety is that products of singular distributions do not generally commute with the weak-limit procedure when their singular supports overlap. As a result, the manipulations leading to Eq.~\eqref{structure2} are rigorously justified only for the regular, state-dependent part of the relevant correlation functions.
 This is sufficient for applications to thermal states, where the finite-temperature contribution to correlation functions is finite and the universal singular vacuum structure has been subtracted.\footnote{States whose two-point functions do not exhibit the correct short-distance Hadamard structure lead to divergences in local observables --- such as the stress-energy tensor --- that cannot be removed by standard renormalization procedures. As a result, such states are regarded as unphysical. Similar issues also appear to constrain the construction of coherent states in QFT, as discussed in Refs.~\cite{Berezhiani:2021gph, Berezhiani:2023uwt}.   }

\section{Dirac delta derivatives and linear response theory}

\label{sec_linear}

The most interesting operator in~\eqref{structure2}  is the one proportional to the derivative of the delta. Roughly speaking,  this term can be interpreted as the first-order correction 
in a Taylor expansion which  displaces the Minkowski lightcone to order $\kappa$, 
\begin{equation} \label{story}
\delta(t^2 - \vec x^{\, 2}) + \kappa\,  t \, {\cal O}( x) \delta'(t^2 - \vec x^{\, 2})  \ \simeq \ \delta(t^2 - \vec x^{\, 2} +  \kappa \,t\,  {\cal O}( x)  ) \, .
\end{equation}

Here, the expression on the RHS should be understood as a formal
all-order resummation. At every finite order in $\kappa$, the perturbative
expansion contains only delta functions and their derivatives, all supported
on the unperturbed Minkowski light cone. A displacement of the support is
therefore invisible at any finite order and can emerge only after resumming
the entire series.\footnote{The elementary example is
$\delta(u-a)\sim\sum_{n=0}^{\infty}(-a)^n\delta^{(n)}(u)/n!$.
Every term in the series is supported at $u=0$, whereas the resummed
distribution is supported at $u=a$. When integrated over a test function
$J$, the partial sums reproduce the Taylor polynomials of $J$ around $u=0$,
which need not converge to $J(a)$ for a generic smooth function. The series
therefore does not converge to the translated delta in the usual topology of
tempered distributions. This is the origin of the analyticity requirement
discussed below.}

One can guess the expression of ${\cal O}(x)$ simply by comparing the above with the lightcone deformation induced by a classical field, eq.~\eqref{3}. Indeed, after some lengthy calculation shown in App.~\ref{A}, one finds 
\begin{equation} \label{o}
{\cal O}(x) =  - \int_0^t dt' \, h_{ij}(t', t' \hat x)\,  \hat x_i \hat x_j  \, ,
\end{equation}
so the support of the delta in~\eqref{story} and eq.~\eqref{3} coincide at the level of expectation values. 

This mechanism, ubiquitous in derivatively coupled theories, has been studied in~\cite{Cintia:2025fzn}. There, we computed the commutator of a \emph{$P(X)$ theory} which displays a propagation speed different than the speed of light on certain Lorentz-breaking backgrounds. In a very similar manner,  a curved spacetime can be seen as Minkowski space with a ``background of gravitons" in the effective theory~\eqref{theory} (see also~\cite{Berezhiani:2021zst,Komissarov:2022gax, Berezhiani:2024boz}). As this theory is Lorentz invariant, the perturbative expansion of the commutator~\eqref{structure2} has a Lorentz-invariant form. The delta derivative terms are responsible for reconciling the Minkowski --- with the geometric curved-spacetime--- perspectives, by acquiring non-vanishing expectation values on Lorentz-breaking field configurations.

In order to give a precise sense to~\eqref{story} one should integrate these terms over a smooth tempered function. 
But this is already implicit in any practical use of the commutator! Most notably, in linear response theory, the smooth tempered function hitting the commutator is just the classical source $J(z)$ causing the perturbation of the field at some event $x$, 
\begin{equation}
    \phi(x) - \phi_{J=0}(x) = i \int d^4 z \, \theta(x^0 - z^0) [\phi(x), \phi(z)] \, J(z) \, .
    \label{eq:linearResponse}
\end{equation}
We assume $J(z)$ to be a smooth real function which is mostly supported in a region of size $R$ centered at the origin of the coordinates.  We want to evaluate the field at positive times with $ x^0 \gg R$, so that, effectively,  $\theta(x^0 - z^0)\simeq1$ always. 
We obtain
\begin{align}
    \phi(x)-\phi_{J=0}(x)= &\ \frac{1}{2\pi} \int d^4z\, \left[\delta(x^0-z^0 - r)-  \frac{\kappa}{2} \mathcal{O}(x,z)\, \partial_{z_0}\delta(x^0-z^0 - r)\right]\frac{J(z^0,\vec{z}\, )}{2r}\nonumber\\[2mm] =& \ \frac{1}{4\pi}\int  d^3z  \, \frac{1}{r} \left[
    J(t_{ret},\vec{z}\, )+\frac{\kappa}{2} \mathcal{O}(x,z)\dot J(t_{ret},\vec{z}\, )+\ \dots \ \right]
    \end{align}
where we have defined $r=|\vec{x}-\vec{z}|$, and $ t_{ret} = x^0-r$. Also, we have omitted terms in which the integration by parts hits the operator $\mathcal{O}(x,z)$, as they only (slightly) modify the intensity of the signal, not its propagation. The expression inside the square brackets in the last integral is simply the current $J$ evaluated at a shifted retarded time,
\begin{equation} \label{eqtrans}
t_{ret} \to t_{ret} + \frac{\kappa}{2} \mathcal{O}(x,z)\, .
\end{equation} 

At any finite order, integration by parts has turned the derivatives of the
delta into ordinary derivatives of the smooth function $J$. Equation~\eqref{eqtrans},
however, amounts to resumming these derivatives into the Taylor series of
$J$. For this series to reconstruct an actual translation, $J$ must be analytic over the relevant range of its argument. This excludes sources of compact support whose boundary lies inside this range. To accommodate this case while retaining the interpretation of the perturbative series as a finite translation, one must work with a class of test functions with weaker localization properties.
For such sufficiently analytic profiles, in the large-distance limit,
$R\ll r$, one could go a step further and approximate
\begin{align}
    \phi(x)-\phi_{J=0}(x)\ \simeq \ \frac{1}{4 \pi r} \
    Q\left(x^0-r+\frac{\kappa}{2}\mathcal{O}(x,0)\right)\, ,
    \label{monopole}
\end{align}
where the charge $Q(t)$ is the spatial integral of $J(t,\vec x)$. Of course,
the displacement becomes appreciable if the typical light-cone shift induced
by the metric field, $\sim\kappa\mathcal{O}$, is comparable to or larger than
the characteristic size $R$ of the source. Since $\mathcal{O}$ is an integral
operator, this is a cumulative effect that depends both on the strength of
the gravitational field and on the total propagation time.

These observations clarify the distinction between the peak and the front of
the signal. At the level of \emph{expectation values}, the perturbative
mechanism under discussion can deform the shape of a signal and, depending on
the sign of $\langle\mathcal{O}(x,z)\rangle$, shift its peak outside the
Minkowski light cone. However, a source of strictly compact support cannot be displaced outside the lightcone at any finite order in perturbation theory by this mechanism. 
In this sense,  Minkowski
causal support is preserved order by order in perturbation theory. A displacement of the \emph{front}---rather than the \emph{peak}---of the signal would be an intrinsically all-order effect, inaccessible to any finite-order calculation.\footnote{We thank
Brando Bellazzini for discussions on this point.}

\section{Evaluating the lightcone uncertainty}

\label{sec_5}

Since we now have the expression of the operator-valued commutator, we now want to take a step beyond its expectation value and evaluate the probability that the commutator be non-vanishing at some point $x$, $ P([\phi(x), \phi(0)] \neq 0)$.
In particular, we will be interested in the gravitons' vacuum and thermal states, both of which are characterized by $\langle {\cal O}(x) \rangle =0$ and thus feature no deviations from the Minkowski lightcone at the background level.\footnote{In the case of a thermal state, deviations of the actual expectation value of the commutator from the Minkowski lightcone happen only at order $\kappa^2$~\cite{CPR}} As the commutator has support at $t^2 - \vec x^{\, 2} +  \kappa \,t\,  {\cal O}( x)=0$ (eq.~\ref{story}), this is tantamount to studying the probability distribution of the operator ${\cal O}( x)$---in other words, the \emph{probability $P(A)$ that the operator ${\cal O}(x)$ evaluates some real number $A$}. 

As demonstrated in App.~\ref{B}, this probability is Gaussian at the lowest order in $\kappa$
\begin{equation} \label{probA}
    P\big(A \big) \ \propto \ \exp\left(-\frac{A^2}{2\langle {\cal O}^2(x) \rangle }\right)\, ,
\end{equation}
so we just need to compute
\begin{equation}
\label{eq:variance}
    \langle \mathcal{O}^2(x)\rangle=\int_0^t dt_1\int_0^t dt_2 \langle h_{ij}(t_1,t_1 \hat{x}) h_{kl}(t_2,t_2 \hat{x})\rangle \hat{x}^i \hat{x}^j \hat{x}^k \hat{x}^l\, .
\end{equation}
The two-point function inside the integral is the Wightman function of free gravitons in $TT-$gauge. In a thermal state of gravitons at temperature $T = \beta^{-1}$, this correlator splits into a vacuum and a purely thermal part,  
\begin{equation}
     \langle h_{ij}(x) h_{kl}(y)\rangle = \int\frac{d^4 k }{(2 \pi)^4} e^{-i k(x-y)} \left[D^{(0)}(k) + D^{(th)}(k)\right]  \left(P_{ik}P_{jl} + P_{il}P_{jk}- P_{ij}P_{lk}\right)\, ,
\end{equation}
with 
{\begin{equation} \label{propa}
D^{(0)} =4  \pi \delta(k^2) \theta(k^0)\, , \qquad  \qquad D^{(th)} = \frac{4 \pi \delta(k^2)}{e^{\beta |k^0|} -1} \, ,
\end{equation}}
and the tensorial structure is the $TT-$projector, with $P_{ij} = \delta_{ij} - \hat k_i \hat k_j$. 

The integral in~\eqref{eq:variance} is computed in App.~\ref{B}. We present the results separately, starting from the purely thermal contribution. 

\subsection{The finite thermal part}

The purely thermal contribution to the integral~\eqref{eq:variance} cannot be evaluated in closed form. However, its large- and short-time behaviours can be extracted. The details of this computation are presented in Appendix~\ref{B}. For the cumulative effects that we are after, we are clearly more interested in the long-time behavior $t\gg T^{-1}$. In this limit we get 
\begin{equation} \label{lineart}
\langle {\cal O}(x)^2 \rangle_\beta = \frac{ 2\, tT}{3\pi} + {\cal O}\Big( \log(tT)\Big).
\end{equation}
As far as we know, this result has never appeared before. The short-time behavior at $t\ll T^{-1}$, on the other hand, has already been estimated by Ford in~\cite{Ford:1994cr}. We further comment on this below in Sec.~\ref{Ford}.

Finally, using Eq.~\eqref{story}, we set $A=\frac{\vec x^{\, 2} -  t^2 }{\kappa t} $ in~\eqref{probA}, and this allows us to translate this result into the probability distribution for the commutator in a thermal state to be non-vanishing
  \begin{equation} \label{probacom}
 P\Big([\phi(x), \phi(0)] \neq 0\Big) \ \propto \ \exp\left(- \frac{3 (\vec x^{\, 2} - t^2)^2}{32\,  G_N t^3  T}\right) \, .
  \end{equation}
  In other words, \emph{$\vec x^{\, 2}$ is a Gaussian variable centered at $t^2$ with variance}

  \begin{equation} \label{variancexsq}
    \text{Var}(\vec{x}^{\, 2}) = \frac{16 \, G_N T t^3}{3}\, .
  \end{equation}

In order to get more intuition of the strength of the effect, it is convenient to phrase it in terms of $\delta x = |\vec x| - t$, i.e., the departure from the classical lightcone. This quantity averages to zero and directly represents the physical distance of the fluctuation from the classical lightcone. One sees  that
\begin{equation} \label{most}
\text{Var}(\delta x) \, = \,  \frac{4 G_N T \, t}{3}\, .
\end{equation}
This result is consistent with a \emph{naive} picture of the fluctuations around the classical lightcone as a random walk\footnote{We thank A. Nicolis and M. Mirbabayi for pointing this out.} in that the spread in $\delta x$ grows as $t^{1/2}$. 
This variance, as well as the vacuum one, is consistent with the
time-delay analysis due to intrinsic graviton fluctuations in interferometers of Ref.~\cite{Freidel-Oberfrank}.

 \subsection{Vacuum contribution}
The above represents the thermal result to which we have subtracted the universal vacuum contribution. We now examine the latter more closely. The correlator  
$D^{(0)}$ in~\eqref{propa} leads to a divergent variance, as its double integration along the same null segment in Eq.~\eqref{eq:variance} is UV divergent at coincidence.
 This reflects the universal short-distance structure of Hadamard states, but also the assumed point-like nature of the source, in the following sense. In linear response theory, the commutator is used to propagate a signal, as shown above in Sec.~\ref{sec_linear}. One way to regulate this divergence is to consider the effects of an \emph{extended source}  of finite size $R$ for such a signal.

Equivalently, and more easily, one can directly introduce in~\eqref{eq:variance}
 a UV short-distance regulator $R$, playing the role of the characteristic length-scale of the source. 
  With this procedure, the vacuum variance at large times turns out to be (see App.~\ref{B} for details)
 \begin{equation}
     \langle \mathcal{O}^2(x)\rangle_0\ \simeq \ \log\left(\frac{t}{R}\right)~.
     \label{vacvariance}
 \end{equation}
which leads to a light-cone variance
\begin{equation} \label{vacuumvar}
    \text{Var}(\vec{x}^{\, 2})\ \simeq \ G_N t^2 \log\left(\frac{t}{R}\right)~.
\end{equation}
Equivalently, 
\begin{equation} \label{vacuumvar}
\text{Var}(\delta x)\ \simeq \ G_N  \log\left(\frac{t}{R}\right)~.
\end{equation}

While a thermal state directly breaks boosts, one might be surprised to find a Lorentz-violating result in the vacuum. Arguably, however, Lorentz invariance \emph{is} broken, by considering a source extended in spacetime and by the gauge choice---that of the set of comoving geodesic observers that define the frame (see  Sec.~\ref{frame}).
 
 The variance~\eqref{vacuumvar} grows \emph{slower} than the thermal one, so vacuum effects become subleading at large distance.
Perhaps more importantly, eq.~\eqref{vacuumvar} is regulator-dependent --- ultimately, source-dependent. Taken at face value, it suggests that a very localized source can induce a large lightcone uncertainty. While such behavior is not inconsistent with quantum intuition, it is at odds with the sharp causal structure normally associated with local quantum field theory in the absence of gravity. This raises the possibility that the source-dependent contribution~\eqref{vacuumvar} is unphysical and should be subtracted, much like the vacuum entanglement entropy is removed when defining the thermal entropy of a spatial region.

\section{Discussion} \label{comments}

In summary, we have computed the variance of the causal lightcone probed by a test scalar field in the presence of gravitational fluctuations about Minkowski spacetime. At leading order in $G_N$, the probability that the commutator is nonzero is Gaussian in the variable $\vec{x}^{\,2}$. The resulting lightcone uncertainty is more transparently expressed in terms of the displacement $\delta x$ from the classical lightcone. For a thermal graviton state, its root-mean-square value grows at late times as $\sqrt{\text{Var}(\delta x)}\sim t^{1/2}$, like in a random walk. This is expressed in eq.~\eqref{most}, which is the main result of this paper. From this result, a universal vacuum contribution~\eqref{vacuumvar} has been subtracted, which, albeit UV divergent, becomes subleading at large separation. 

We conclude with a list of comments, ordered roughly from the more technical to the more speculative.

\subsection{ Comparison with the results of L. H. Ford} \label{Ford}

While finalizing our draft, we became aware of the important work by Laurence H. Ford and collaborators on fluctuations of causality on general gravitational states~\cite{Ford:1994cr,Ford:1996qc,Yu:1999pq,Yu:1999wg,Yu:2000yf}. We highlight the main differences and improvements between our approach and these earlier works.

In Ref.~\cite{Ford:1994cr}, Ford postulates that, near the light cone, the retarded propagator of a scalar field in the presence of quantum gravity retains the form of a delta function supported on the modified operator-valued light cone  $\vec x^{\, 2} = t^2 + \kappa t {\cal O}(x)$ (in his notation, $\kappa t {\cal O}(x) = \sigma_1$ and he uses units with $\kappa =1/2$). In this work, we have proved that this ansatz is correct and derived the commutator from a first-principles computation using the interaction-picture formalism. In particular, we derived this result starting from a theory in which gravity is treated as a quantum field from the outset, without relying on results from quantum field theory in curved spacetime as in Ford's computation.

Moreover, Ford evaluates \(\langle\mathcal{O}(x)^2\rangle\), including in a thermal graviton state, using a large-wavelength ``local" approximation (see Eqs.~(49) and (50) of Ref.~\cite{Ford:1994cr}). However, this approximation captures only the regime \(t\lesssim T^{-1}\). Indeed, up to a numerical coefficient, his result is consistent with the short-time behavior that we find  in App.~\ref{B}, \emph{i.e.}
  \begin{equation} \nonumber
 \qquad\text{Var}(\vec{x}^{\, 2}) \simeq  G_N T^2 t^4   \qquad \qquad  \text{ ( \(t\lesssim T^{-1}\) limit)}\, .
  \end{equation}
However, this range of lengths is typically microscopic. Within its regime of validity, the above variance is bounded by $\text{Var}(\vec{x}^{\, 2}) \lesssim G_N T^{-2}$, and just represents a short microscopic transient before the onset of the cumulative effect discussed in this paper, which clearly dominates at large times. At  \(t\gg T^{-1}\),  the nonlocal nature of \(\mathcal{O}(x)\) cannot be neglected and yields the asymptotic behavior~\eqref{variancexsq}. 

Finally, while in \cite{Ford:1994cr} the vacuum contribution to the variance was simply subtracted from $\langle \mathcal{O}^2(x)\rangle$, in this work we have evaluated it explicitly. In particular, we have regularized its divergence and interpreted the resulting expression within the framework of linear-response theory.

\subsection{Consistency of the setup} \label{setup}

One might correctly object that a thermal state of gravitons is incompatible with Minkowski spacetime once backreaction is taken into account. Indeed, a thermal energy density of $\sim T^4$ produces a curvature radius of $H^{-1} \sim 1/(\kappa T^2)$. This is the age of a Universe dominated by radiation at that temperature and also, strictly speaking, the maximum spatial size of our setup. It is tempting to speculate, however, that the broader message of our result~\eqref{lineart} is a secular linear growth in time of $\langle {\cal O}(x)^2 \rangle_\beta$, with a coefficient proportional to the temperature, and that this could be essentially extrapolated to more realistic scenarios, e.g., when the temperature is smoothly changing in time. 

A specular question is the role of higher $G_N$-corrections   to the expression of the variance, eq.~\eqref{variancexsq}. In order to understand the power counting, it is useful to focus on the dimensionless quantity $ \text{Var}(\delta x)/t^2$. According to the leading order result~\eqref{variancexsq} this is proportional to the effective gravitational coupling at temperature $T$, $G_N T^2$, divided by the combination $T t$, which is the number of thermal correlation lenghts in a time $t$.  Presumably, higher terms in the EFT expansion just correspond to higher powers of the EFT coupling, giving a series of the type
\begin{equation}
\frac{\text{Var}(\delta x)}{t^2} = \frac{1}{T t}\left[ \frac{4}{3} G_N T^2 + c_1 \left(G_N T^2\right)^2 + \dots\right]\, .
\end{equation}
The first term in square brackets matches the result~\eqref{most}, while the numerical coefficients  $c_1$ etc., are yet to be evaluated. In summary, one should expect the next-to-leading term to be proportional to $G_N^2 T^3$ with the same $t$-dependence as the leading term.  This is consistent with our result representing the leading contribution at large times and low energy. Terms dying out more quickly at large times (\emph{e.g.}, $\sim 1/(Tt)^2$) seem also possible, in principle.

\subsection{ Black holes }

The uncertainty expressed in eq.~\eqref{most} is exceedingly small under ordinary conditions. At room temperature, the time required for the lightcone spread to reach the meter scale vastly exceeds the age of the universe. In flat spacetime the effect is particularly inconsequential because the lightcone grows much faster than its spread. 
However, it could be of relevance in situations when the lightcone ``stops growing". The latter is a loose characterization of a black hole horizon, which can be thought of as a bundle of light rays with vanishing expansion. Incidentally, this is also where an approximately thermal spectrum of gravitons can be found, in the form of  Hawking radiation. 

This suggests the naive  extrapolation of eq.~\eqref{most} to the Hawking temperature, $T\sim R_s^{-1}$, where $R_s$ is the black hole radius. With this substitution, the lightcone uncertainty becomes comparable to the size of the black hole $R_s$ after a time
\begin{equation}
    t\sim T^{-1}S\,,
\end{equation}
where $S\sim 1/(G_N T^2)$ is the black hole entropy. This is parametrically the evaporation/Page time, at which information is expected to begin emerging in the Hawking radiation~\cite{Page:1993df}.

Of course, the application of our formalism to black holes will require a separate study. 
Still, the cumulative effects found in this paper are clearly perturbative in $G_N$, which seems in contrast with the general expectation that deviations from semiclassical gravity should be of ${\cal O}(e^{-S})$. On this point, it is worth citing the work of Dvali and collaborators which, in a series of papers (\emph{e.g.}~\cite{Dvali:2013eja, Dvali:2017ruz, Dvali:2017eba} and references therein) argue for a \emph{quantum break-time} for black hole.
According to these authors, black hole dynamics is  dominated before the Page time by cumulative quantum effects which are perturbative in $G_N$~\cite{Dvali:2011aa, Dvali:2012en,Dvali:2015aja, Dvali:2020wft}. Limits on the semiclassical approximation due to the variance of the energy momentum tensor have also been discussed in~\cite{Perez:2025tvg}.

The black hole information paradox is routinely formulated and refined by setting  \emph{gedanken} ``Alice and Bob" experiments on a perfectly classical background spacetime, with a sharp horizon and exact causal structure (e.g.~\cite{Susskind:1993mu,Sekino:2008he,Almheiri:2012rt}). Our results may suggest that aspects of this classical picture could instead gradually cease to apply. 
One might be tempted to associate any breakdown of classical spacetime with a sharp, localized phenomenon, like in the famously debated firewall scenario~\cite{Almheiri:2012rt}. Here, on the opposite, the classical spacetime picture appears to fail in the same way that flat spacetime fails in general relativity itself---not abruptly, but at sufficiently large mutual separations~\cite{Piazza:2021ojr,Piazza:2022amf}.

\subsection*{Acknowledgments}
 We thank  Gia Dvali, Philipp Hoenn, Christian Marinoni, Alberto Nicolis, Francesco Nitti, Alejandro Perez, Alessandro Podo, Riccardo Rattazzi, Marcello Romano and especially Brando Bellazzini and Mehrdad Mirbabayi for discussions and comments on the draft.
This work received support from the French government under the France 2030 investment plan, as part of the Initiative d'Excellence d'Aix-Marseille Universit\'e - A*MIDEX (AMX-19-IET-012). It was also supported by the ``action th\'ematique" Cosmology-Galaxies (ATCG) of the CNRS/INSU PN Astro and by the {\it Agence Nationale de la Recherche} under the grant ANR-24-CE31-6963-01.

\appendix

\section{Derivation of the operatorial commmutator }\label{app:commutator} \label{A}
In this Appendix we present the derivation of the interacting commutator
\begin{align}  \nonumber
2 \pi i \, [\phi(x), \phi(0)] & \ =\ \mathrm{sign}(t)\Big[\delta(t^2 - \vec x^{\, 2}) \\ & \ \, + \, \kappa\,  \Big( t \, {\cal O}( x) \delta'(t^2 - \vec x^{\, 2}) + {\cal O}_1( x) \delta(t^2 - \vec x^{\, 2}) + {\cal O}_2( x) \theta(t^2 - \vec x^{\, 2}) \Big)\,\Big] . 
\label{eqapp:comm}
\end{align}
which generalizes Eq.~\eqref{structure2} without assuming a definite sign for $t$.

The first-order correction to the free commutator is obtained from the Dyson series, and is provided in Eq.~\eqref{eq:comm-intpic}. After repeated integration by parts, this expression can be recast as
\begin{align}
     [\phi(x),\phi(y)] &= i\Delta(x-y)-i\kappa \partial_{x^i}\partial_{x^j}\int^{t_x}_{t_y}\dd[4] z\, h^{ij}(z)\Delta(z-x)\Delta(z-y) \nonumber\\
     &= i\Delta(x-y)-i\kappa \partial_{x^i}\partial_{x^j}I^{ij}(x,y)~.
     \label{eqapp:comm-Heispic}
\end{align}
In what follows, we set $y=0$, write $x=(t,\vec x)$ and denote $r\equiv|\vec x|$.  

Let us start by assuming $t>0$, and check the case, and come back to $t<0$ later.
Assuming that the field $h^{ij}$ is regular on all spacelike hypersurfaces between $0$ and $t$, we integrate the above expression over $z^0$.  
We find
\begin{align}
    I^{ij}(x) &=-\frac{1}{4\pi^2} \int\mathrm{d}^3 \vec{z} \int^{t}_{0}\dd z^0\,  h^{ij}(z^0,\vec z)\delta((z-x)^2)\delta(z^2)\nonumber\\
    &=-\frac{1}{4\pi^2} \int\mathrm{d}^3 \vec{z}\, \,\theta(t-|\vec z|) \theta(|\vec z|) \delta\left((|\vec z| -t)^2-(\vec z -\vec x)^2\right)\frac{ h^{ij}(|\vec z|,\vec z)}{2|\vec z|}.
\end{align}
where we have an overall minus from the $\sign(z^0-t)$ function. The time-integral has been performed using the positive root of the $\delta(z^2)$, which is the only one supported on the region $0<z^0<t$. We also write explicitly Heaviside functions to keep track of the integration domain. 

We then introduce the angle $\vartheta$ between the vector $\vec x$ and $\vec z$. The dirac delta reads
\begin{flalign}
    \int^{1}_{-1} d\cos\vartheta \,\delta(2(r\cos \vartheta-t)&|\vec z|-(r^2-t^2))\theta(t-|\vec z|) \theta(|\vec z|)\\&=\frac{1}{2r|z|} \left\{\theta(|\vec{z}|-\frac{(t-r)}{2})-\theta(|\vec{z}|-\frac{(t+r)}{2})\right\}\theta(t-r)
\end{flalign}
To derive the above relation, it is convenient to express the integration domain in $\cos\vartheta$ using Heaviside step functions $\theta(\cos\vartheta+1)-\theta(\cos\vartheta-1)$. One can then perform the integration over $\cos\vartheta$ via the delta function, and subsequently rewrite the resulting expression in terms of Heaviside functions as above. In this procedure, we assume $t>r$ since the alternative case yields a vanishing contribution when combined with $\theta(t-|\vec z|) \theta(|\vec z|)$.

 We get
 \begin{align}
    I^{ij}(x) =-\frac{1}{8
    \pi r} \theta(t-r)\int \frac{d\varphi}{2\pi}\int_{\frac{t-r}{2}}^{\frac{t+r}{2}} d\tau \,h^{ij}(\tau,\tau\,\hat{n}(\varphi,\tau)), \quad \text{with}\quad t>0.
\end{align}
Here, we made the relabelling $\tau=|\vec{z}|$.

The unit vector $\hat n(\tau,\varphi)$ takes the standard spherical form, which in a Cartesian frame can be written as
\begin{align}
    &\hat n(\sigma,\varphi) = \left(
        \begin{array}{c}
        \sin\vartheta(\tau)\cos\varphi \\
      \sin\vartheta(\tau)\sin\varphi \\
        \cos\vartheta(\tau)
        \end{array}
    \right),
    \quad \text{where} \quad 
   \cos{\vartheta(\tau)}=\frac{r^2-t^2}{2r\tau}+\frac{t}{r}.
\end{align}
Here, $\vartheta(\tau)$ is determined by the solution of the delta-function constraint.

In contrast, assuming $t<0$ would end up replacing $-\theta(t-r)$ with $\theta(-t-r)$. Combining the two cases, the expression which is valid for all $t$ reads
\begin{align}
    I^{ij}(x) =-\frac{\text{sign}(t)}{8
    \pi r} \theta(t^2-r^2)\int \frac{d\varphi}{2\pi}\int_{\frac{t-r}{2}}^{\frac{t+r}{2}} {d\tau} \,h^{ij}(\tau,\tau\,\hat{n}(\tau,\varphi))~.
    \label{eq:mi-app}
\end{align}

We finish by deriving $\partial_{x^i}\partial_{x^j}I^{ij}$. Since Eq.~\eqref{eq:mi-app} depends only on the radial distance $r$, the second spatial derivative reads
\begin{align}
 \partial_{x^i}\partial_{x^j}I^{ij}&=\left(\hat x^i \hat x^j \partial_r^2 +\frac{\delta_{ij}-\hat x^i\hat x^j}{r}\partial_r\right) I^{ij}.
    \label{eq:Hessian-mi-app}
\end{align}
To simplify the derivation, we introduce the field $H^{ij}(t,r)$ as follows
\begin{align}
    H^{ij}(t,r)&= \int^{2\pi}_0\frac{\dd\varphi}{2\pi} \int_{\frac{t-r}{2}}^{\frac{t+r}{2}} \dd \tau\, h^{ij}\left(\tau,\tau\hat n(\tau,\varphi)\right).
\end{align}
Then, the first derivative and second radial derivative of $I^{ij}$ read
\begin{align}
    \partial_r I^{ij} &= \frac{\sign t}{8\pi}\left[
    2\delta(t^2-r^2)H^{ij}+\theta(t^2-r^2)\left(\frac{1}{r^2}H^{ij}-\frac{1}{r}\partial_r H^{ij}\right)\right] \\
    \partial_r^2 I^{ij} &= -\frac{\sign t}{8\pi}\bigg[4r\delta'(t^2-r^2)H^{ij}+\delta(t^2-r^2)\left(\frac{2}{r}H^{ij}-4\partial_r H^{ij}\right) \nonumber\\
    &\qquad\qquad\qquad+\theta(t^2-r^2)\left(\frac{2}{r^3}H^{ij}-\frac{2}{r^2}\partial_r H^{ij}+\frac{1}{r}\partial_r^2 H^{ij}\right)\bigg].
\end{align}
Combining everything into \eqref{eq:Hessian-mi-app}, we finally have derived
the linear order correction to the commutator of a massless scalar field coupled to a graviton in \textit{TT}-gauge that reads
\begin{flalign}
    2\pi i[\phi(x),\phi(0)] = \sign(t)\delta(t^2-\vec x^2)-&\kappa\sign(t) \hat x_i \hat x_j \bigg[ r\delta'(t^2-\vec x^2)H^{ij} +\frac{\delta(t^2-r^2)}{r}\left(H^{ij}-r\partial_r H^{ij}\right) \nonumber\\
    &+\frac{1}{4r^3}\theta(t^2-\vec x^2)\left(3H^{ij}-3r\partial_r H^{ij} +r^2\partial_r^2 H^{ij}\right)\bigg].
    \label{eq:commutator}
\end{flalign}
Notice that the terms involving $\delta_{ij}$ vanish due to the traceless condition.  The above expression can then be identified with the expansion of the commutator in Eq.~\eqref{eqapp:comm}. Focusing on the coefficient of the derivative of the delta function and evaluating it on its future support, we obtain
\begin{equation}
    \mathcal{O}(x)=- \int_{0}^{t} \dd \tau\, h^{ij}\left(\tau,\tau\, \hat{x}\right)\hat x_i \hat x_j.
\end{equation}
Here, the $\varphi$ integral has been carried out, since on the light cone $\hat n(\sigma,\varphi)$ reduces to $\sign(t)\hat x$. As we see, it matches \eqref{o} we provided in the main text.

 \section{Probability distribution and variance}
  \label{B}
In this Appendix, we show that, to leading order in $\kappa$, the probability distribution associated with the operator $\mathcal{O}(x)$ is Gaussian. We the proceed to calculate $\langle \mathcal{O}^2(x)\rangle$ in a thermal state and in the vacuum. 

Probabilities in quantum mechanics are expectation values of projection
operators. In the present case, the projector onto a given value $A$ of $\mathcal{O}(x)$ is represented by the one-dimensional delta function $\delta(\mathcal{O}(x)-A)$. The corresponding probability is therefore
\begin{equation}
    P\big(A \big) =\langle\delta(\mathcal{O}(x)-A) \rangle\ =\int_{-\infty}^\infty\frac{dp}{(2\pi)} \langle e^{-ip \mathcal{O}(x)}\rangle e^{ip A}
    \label{eq:pintegral}
\end{equation}
To evaluate the characteristic function $\langle e^{-ip\mathcal{O}(x)}\rangle$, we expand the exponential in powers of $\mathcal{O}(x)$. At leading order in $\kappa$, the graviton theory is Gaussian and Wick's theorem implies that all odd moments vanish. Hence, we only need to keep track on the even ones, finding
\begin{equation}
    \langle e^{ip \mathcal{O}(x)}\rangle=\sum_{n=0}^\infty\frac{(-ip)^{2n}}{2n!}\langle\mathcal{O}^{2n}(x)\rangle=\sum_{n=0}^\infty\frac{(-p^2)^{n}}{2^{n}n!}\langle\mathcal{O}^2(x)\rangle^n=e^{-\frac{p^2}{2}\langle\mathcal{O}^2(x)\rangle}
\end{equation}
Substituting this result into the Fourier representation~\eqref{eq:pintegral}, one immediately gets a Gaussian probability distribution~\eqref{probA} after integrating over $p$.

Finally, we derive the vacuum and thermal contributions to the variance of the operator $\mathcal{O}(x)$, given in Eqs.~\eqref{vacvariance} and \eqref{lineart}, respectively. We show in particular that the vacuum variance is divergent and therefore requires the introduction of a length regulator $R$.

The expectation value of the $\langle \mathcal{O}^2(x)\rangle$ reads
\begin{equation}
\label{eq:varianceApp}
    \langle \mathcal{O}^2(x)\rangle=\int_0^t dt_1\int_0^t dt_2 \langle h_{ij}(t_1,t_1 \hat{x}) h_{kl}(t_2,t_2 \hat{x})\rangle \hat{x}^i \hat{x}^j \hat{x}^k \hat{x}^l~\,.
\end{equation}
Since we are dealing with a thermal configuration, the averaging procedure should be understood as the trace of the enclosed operator over a thermal density matrix, with the two-point function provided by the sum of the expressions in Eq.~\eqref{propa}. 

We use this to evaluate the variance. Concerning the vacuum part, we have
\begin{equation}
    \langle h_{ij}(x_1) h_{kl}(x_2)\rangle_0 =2\int\frac{d^4k}{(2 \pi)^3}  \delta(k_\mu k^\mu) \theta(k_0) e^{i k_\mu(x_1-x_2)^\mu}\left(P_{ik}P_{jl} + P_{il}P_{jk}- P_{ij}P_{lk}\right)~,
\end{equation}
with $P_{ij} = \delta_{ij} - \hat k_i \hat k_j$.
Putting all together, we find
\begin{flalign}
    \langle \mathcal{O}^2(x)\rangle_0&=\nonumber\int_0^{R^{-1}}\frac{k^2dk}{2\pi^2}\int_0^t dt_1\int_0^t dt_2 \int_{-1}^1 d\cos\vartheta\int_{0}^\infty dk^0\delta(k_\mu k^\mu)e^{-i (k^0-k\cos\vartheta)(t_1-t_2)}(1-\cos^2\vartheta)^2\\&\approx\frac{4 }{3 \pi ^2}\log \left(\frac{ t}{R}\right),
\end{flalign}
where $\vartheta$ is the angle between the integration variable $\vec k$ and $\vec x$,  and we have used the short notation $k \equiv |\vec k|$. Above, we exploited the following relations
\begin{equation} \label{polar}
 \left(P_{ik}P_{jl} + P_{il}P_{jk}- P_{ij}P_{lk}\right)  \hat x_i \hat x_j  \hat x_k \hat x_l = (1 - \cos\vartheta^2)^2\, ,
 \end{equation}
 as well as the identity
\begin{equation} 
\delta(k_\mu k^\mu) = \frac{\delta(k^0 - k)}{2 k} + \frac{\delta(k^0 + k)}{2 k}~.
\end{equation}
 Moreover, in the last step we considered the limit $t/R\gg1$. As discussed, we had to introduce a momentum cutoff. This can be interpreted as deriving the variance from linear response theory, and setting an effective size of a source that smears the variance.

In contrast, the thermal part of the variance is finite. In this case, the quadratic correlation function we need to integrate over reads
\begin{equation}
    \langle h_{ij}(x_1) h_{kl}(x_2)\rangle_\beta =\int\frac{d^4k}{(2 \pi)^3}  \frac{2\,\delta(k_\mu k^\mu)}{e^{\beta |k^0|}-1}  e^{i k_\mu(x_1-x_2)^\mu}\left(P_{ik}P_{jl} + P_{il}P_{jk}- P_{ij}P_{lk}\right)~.
\end{equation}
The variance reads
\begin{align}
\langle O^2(x)\rangle_\beta = \int_{-\infty}^{+\infty} \!\!\frac{dk^0}{2 \pi^2} \int_{0}^{\infty} \!\!dk k^2 \int_{-1}^1 \!\!d \cos \vartheta \int_0^t \!\!d t_1  \int_0^t\!\! dt_2 \ \frac{\delta(k_\mu k^\mu) e^{i(k^0 - k \cos \vartheta)(t_1-t_2)}}{e^{\beta |k^0|} -1}(1 - \cos\vartheta^2)^2\, ,
\end{align}
Using the same identities exploited for the vacuum part, we integrate over all variables but $k$, obtaining
\begin{equation} \label{56}
 \langle O^2(x)\rangle_\beta =  \frac{2}{3 \pi^2}  \int_{0}^{\infty} \! \frac{dk }{e^{ \beta k} -1}\  \frac{3 \sin(2 k t) + 4 k^3 t^3 - 6 k t}{k^4 t^3}\, .
 \end{equation}
Despite the appearances, this integral is finite because the small-$k$ behavior of the numerator takes care of the apparent IR divergence. 

 We have not been able to solve this integral in closed form, but it is relatively easy to isolate the leading large-time and short-time contributions. 
 By using the dimensionless integration variable $y = kt $ and expanding the Boltzman exponential at $\beta/t \ll 1$ we obtain
 \begin{equation} \label{57app}
 \langle O^2(x)\rangle_\beta\ \approx \ \, \frac{2}{3 \pi^2} \frac{t}{\beta} \int_{0}^{\infty} \! dy  \ \frac{3 \sin(2 y) + 4 y^3 - 6 y}{y^5 } \ = \ \, \frac{2}{3\pi}{t \, T} .
 \end{equation}
In contrast, if $\beta/t \gg 1$, using the integration variable $z=\beta k$, we have
{ \begin{equation} \label{58app}
  \langle O^2(x)\rangle_\beta \ \approx \frac{2t}{3 \pi^2 \beta}\int_{0}^{\infty} \frac{dz}{(e^{z}-1)}  \ \frac{4tz}{5\beta} = \frac4{45} t^2 T^2.
 \end{equation}}
Notice that, in order to obtain these results, the tensorial nature of the graviton has played a significant role. Without summing over the polarizations as in~\eqref{polar}---\textit{i.e.}, had we done the calculation for a scalar---the step corresponding to going from eq.~\eqref{56} to eq.~\eqref{57app} 
should be taken with more care. The resulting late-time behavior in this case is enhanced by a logarithmic factor, \textit{i.e.}, $ \langle {\cal O}(x)^2 \rangle_\beta \sim t T \log(tT)$.

\bibliographystyle{JHEP}
\bibliography{refs.bib}

\end{document}